\documentclass{article}
\usepackage{spconf,amsmath,graphicx}
\usepackage{amsfonts,amsmath,amssymb,bbm}
\usepackage{graphicx,cases,multirow,subfigure}
\usepackage{algpseudocode,algorithmicx,algorithm}
\usepackage[urlcolor=blue,linkcolor=red, citecolor=blue, colorlinks=true]{hyperref}
\usepackage[sort,compress]{cite}
\def\sm{\setminus}
\def\cU{\mathcal{U}} 
\def\cM{\mathcal{M}}
\def\cH{\mathcal{H}}

\def\cR{\mathcal{R}}
\def\cP{\mathcal{P}}
\def\cI{\mathcal I}
\def\cL{\mathcal{L}}
\def\cT{\mathcal T}
\def\cS{\mathcal{S}}
\def\cC{\mathcal{C}}

\def\cD{\mathcal{D}}
\def\bR{\mathbb{R}}
\def\bI{\mathbb{I}}
\def\kq{\mathfrak q}
\def\x{\mathtt x}
\def\s{\mathtt s}
\def\fu{\mathbf u}

\def\kc{\mathfrak C}
\def\0{\mathbf 0}

\DeclareMathOperator\Lips{Lips}
\newtheorem{definition}{Definition}

\newlength\savewidth
\newcommand\shline{\noalign{\global\savewidth\arrayrulewidth\global\arrayrulewidth 1.0pt}\hline\noalign{\global\arrayrulewidth\savewidth}}

\newlength\savedwidth

\title{A New Coherence-Penalized Minimal Path Model with Application to Retinal Vessel Centerline Delineation}
\name{Da Chen\sthanks{chenda@ceremade.dauphine.fr}~and~Laurent D. Cohen}
\address{University Paris Dauphine, PSL Research University\\
             CNRS, UMR 7534, CEREMADE, 75016 Paris, France}

\begin{document}
%
\maketitle
\begin{abstract}
In this paper, we propose a new minimal path model for minimally interactive retinal vessel centerline extraction.   The main contribution lies at the construction of a novel coherence-penalized Riemannian metric in a lifted space, dependently of  the local geometry of tubularity and an external scalar-valued reference feature map.  The globally minimizing curves associated to the proposed metric favour to  pass through a set of retinal vessel segments with low variations of the  feature map, thus can avoid the short branches combination problem and shortcut problem, commonly suffered by the existing minimal path models in the application of retinal imaging. We validate our model on  a series of retinal vessel patches obtained from the  DRIVE and IOSTAR datasets,  showing that our model indeed get promising results.
\end{abstract}
\begin{keywords}
Retinal Vessel Extraction, Minimal Path, Fast Marching Method, Feature Coherence.
\end{keywords}
\section{Introduction}
\label{sec:intro}
The minimal path model is  a powerful tool for  vessel centerline delineation thanks to its  numerical efficiency, global optimality and  elegant mathematical background. In the original formulation~\cite{cohen1997global}, a vessel can be  modelled as a globally minimizing  curve dependent of a metric, through which  the minimal paths can be obtained through the solution to the respective Eikonal partial differential equation (PDE) by the fast marching methods~\cite{sethian1999fast,mirebeau2014anisotropic}.

The classical Cohen-Kimmel model~\cite{cohen1997global} provides a general Eikonal PDE framework for vessel segmentation. One research line  on the improvements of the Cohen-Kimmel model  is to design proper metrics to address different situations. In practice, it is interesting to take into account the orientations which a vessel should have  for minimal path enhancement. This can be done by invoking  anisotropic Riemannian metrics~\cite{benmansour2011tubular} or orientation lifted isotropic metrics~\cite{pechaud2009extraction} for geodesic distance computation.    In order to obtain a smooth minimal path, the sub-Riemannian metric~\cite{bekkers2015pde} and the Finsler elastica metric~\cite{chen2017global} were developed  in an orientation space based on the Eikonal PDE framework, which  in some extent  can find solutions for artery extraction, but also suffer from the short branches combination problem especially when extracting  a long artery. 
Other interesting shortest path models include~\cite{liao2017progressive,wang2013interactive,ulen2015shortest}. 

\begin{figure*}[t]
\centering
\includegraphics[width=5.5cm]{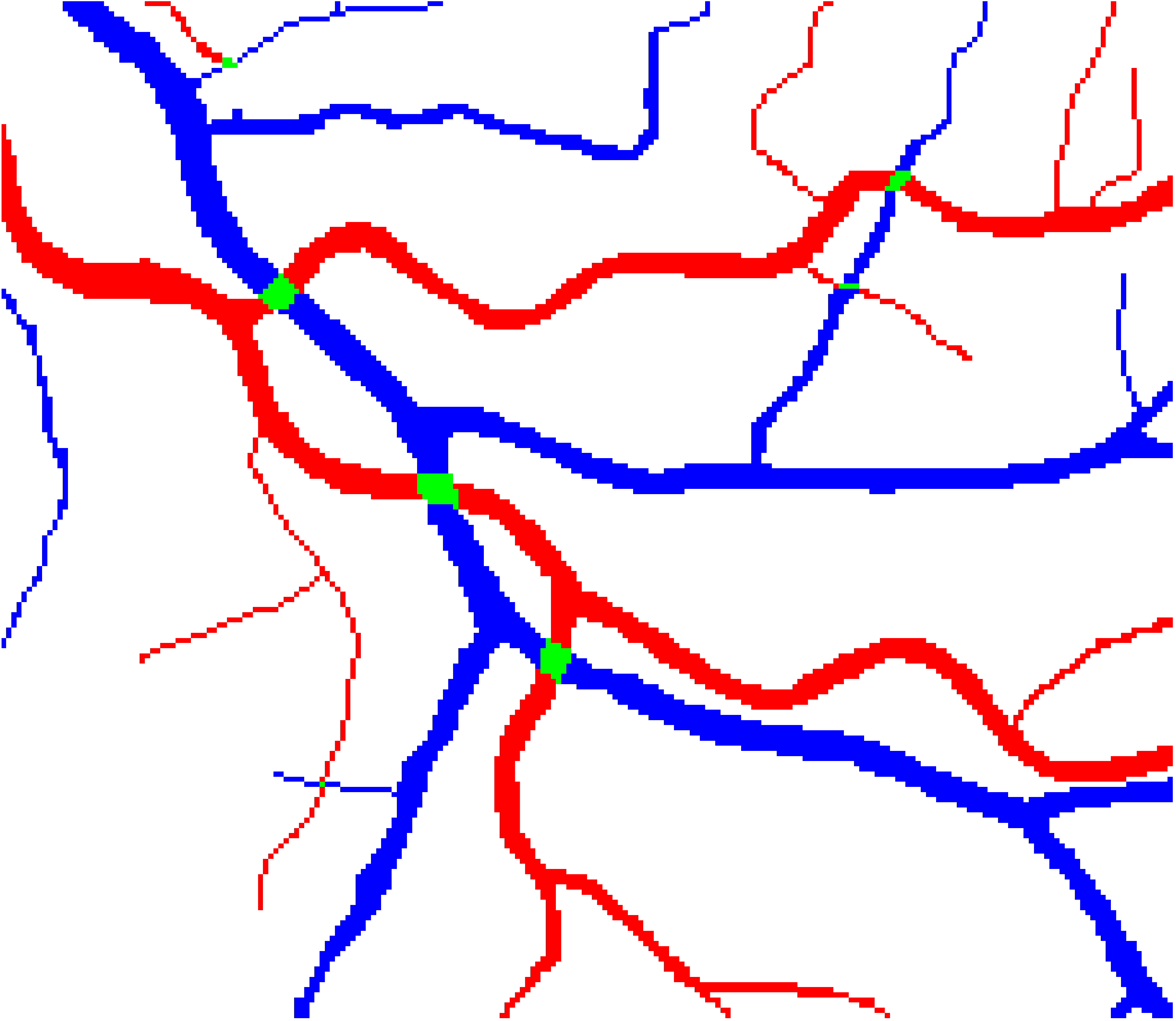}	
\includegraphics[width=5.5cm]{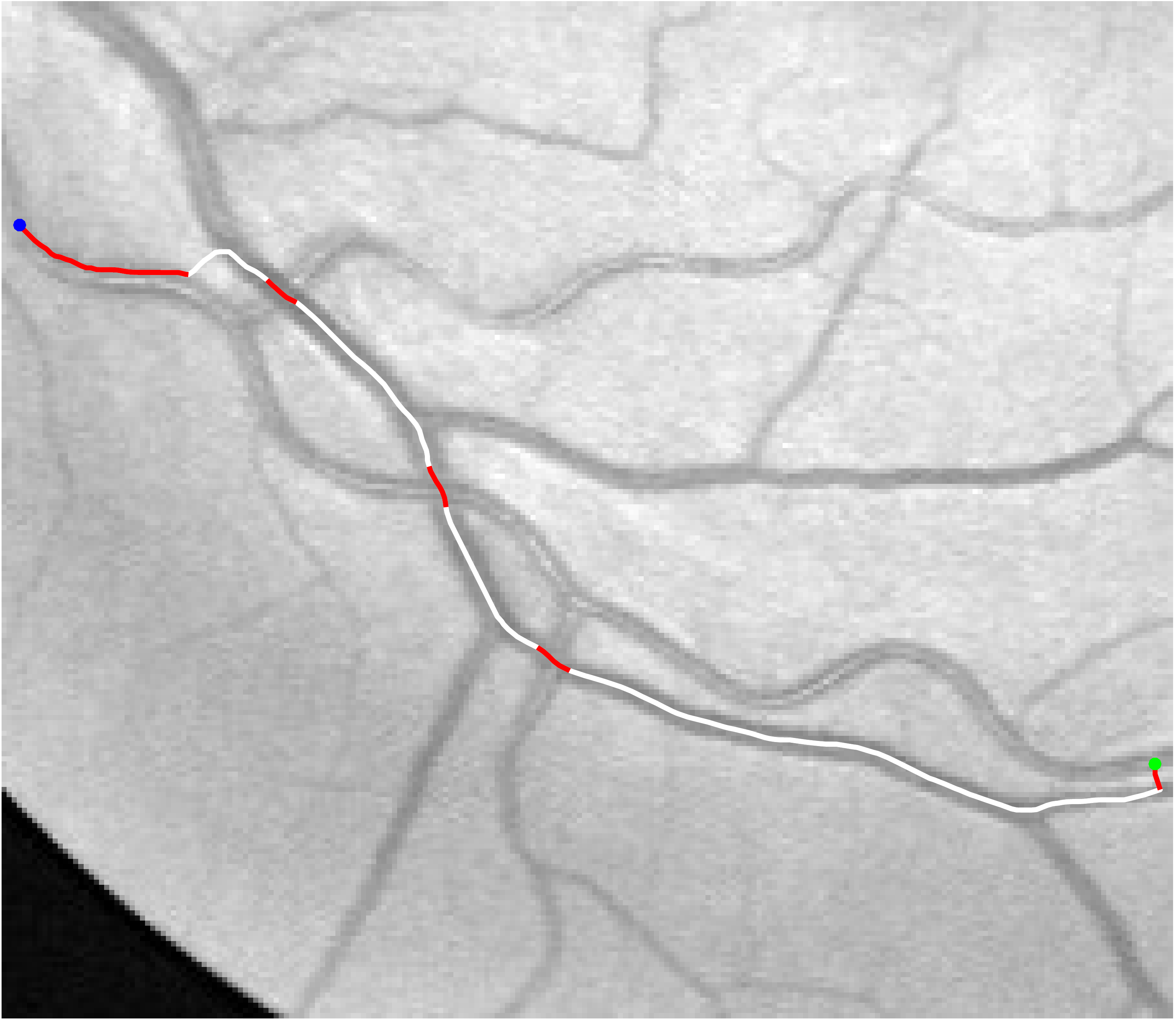}	
\includegraphics[width=5.5cm]{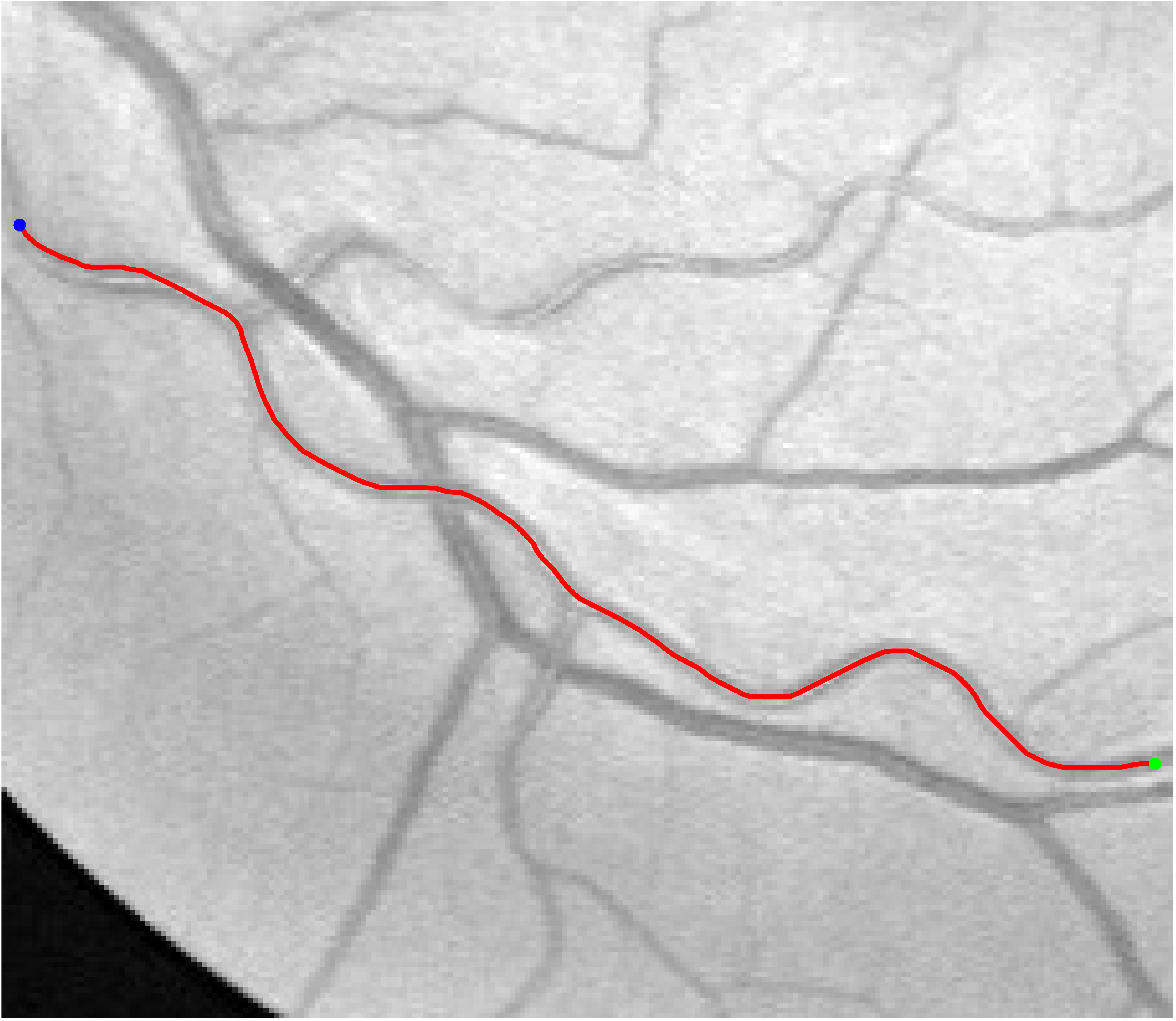}	
\caption{Short branches combination. \textbf{Column 1} Red, blue and green regions respectively indicate arteries, veins and their overlaps. \textbf{Columns 2}-\textbf{3} Results from~\cite{benmansour2011tubular}  and our method.}
\label{fig:ShortBC}
\end{figure*}

In this paper, we propose a coherence-penalized minimal path model, where the associated minimal paths favour to pass by a vessel that is located in the  flatten region of an external  feature map. We observe that along a piece of retinal vessel, the values of gray levels vary slowly. More specifically,  retinal arteries have lower  contrast of gray levels than veins due to the blood materials and imaging modality. In other words, in some extent  the arteries and veins are distinguishable in terms of vesselness values.  Such an observation can be used to solve the  short branches problem  that  the minimal paths associated to a metric may pass through segments belonging to different vessels as shown in Figs.~\ref{fig:ShortBC}b and \ref{fig:ShortBC}c. Fig.~\ref{fig:ShortBC}d shows the result from the proposed method, which can avoid such problem. Fig.~\ref{fig:ShortBC}a gives the artery-vein (AV) groundtruth. In this paper, we denote by blue and green dots the source and end points respectively.

\section{Background on Riemannian Geodesics}
\begin{definition}
Let $\Omega\subset\bR^d$ be an open bounded domain with $d$ the dimension and let $S^+_d$ be the set of positive symmetric definite  (PSD) matrices.  We denote by $\Lips([0,1],\Omega)$ the set of all Lipschitz paths $\eta:[0,1]\to\Omega$. 	
\end{definition}
\begin{definition}
A Riemannian metric $\cP:\Omega\times\bR^d\to[0,\infty]$ is a convex function  such that 
\begin{equation}
\label{eq:RiemannianMetric}	
\cP(x,\fu):=\sqrt{\langle\fu,\cM(x)\,\fu\rangle}=\|\fu\|_{\cM(x)},
\end{equation}
where $\langle\cdot,\cdot\rangle$ denotes the Euclidean scalar product on $\bR^d$ and $\cM:\Omega\to S^+_d$ is a tensor field. 
\end{definition}
A Riemannian geodesic  is  a curve that globally minimizes a geodesic energy $\ell:\Lips([0,1],\Omega)\to [0,\infty]$ associated to a Riemannian metric $\cP$
\begin{equation*}
\ell(\eta)=\int_0^1\cP\big(\eta(t),\dot\eta(t)\big) dt=\int_0^1\sqrt{\langle\dot\eta(t),\cM(\eta(t))\,\dot\eta(t)\rangle}dt,
\end{equation*}
where $\eta\in\Lips([0,1],\Omega)$ and $\dot\eta$ denotes the first-order derivative of $\eta$. The tensor field $\cM$ can be  decomposed  by its eigenvalues $\xi_n$ and eigenvectors $\kq_n$ such that  
\begin{equation}
\label{eq:Tensor}
\cM(x):=\sum_n\xi_n(x) \kq_n(x)\kq_n^{\rm T}(x).	
\end{equation}
For a  source point $s$, the global minimum  of $\ell$ between any point $x\in\Omega$ can be characterized by  the  minimal action map 
\begin{equation*}
\cU_s(x):=\inf\,\{\ell(\eta);\eta\in\Lips([0,1],\Omega),\eta(0)=s,\eta(1)=x\},
\end{equation*}
which is the unique viscosity  solution to  the Eikonal PDE
\begin{equation}
\label{eq:EikonalPDE}
\|\nabla\cU(x)\|_{\cM^{-1}(x)}=1,\, \forall x\in\Omega\sm \{s\},~\text{and}\quad\cU_s(s)=0.
\end{equation}
The geodesic $\cC_{\s,\x}\in\Lips([0,1],\Omega)$ linking  from  $s$ to  $x$ can be computed by re-parameterizing and reversing the geodesic $\rho_{\x,\s}$  by solving the ordinary differential equation (ODE)
\begin{equation}
\label{eq:ODE}	
\begin{cases}
&\rho_{\x,\s}(t)=-\frac{\cM^{-1}(\dot\rho_{\x,\s}(t))\,\nabla\cU(\rho_{\x,\s}(t))}{\|\cM^{-1}(\dot\rho_{\x,\s}(t))\,\nabla\cU(\rho_{\x,\s}(t))\|},\\
&\rho_{\x,\s}(0)=x.	
\end{cases}
\end{equation}

For the task of retinal  vessel centerline extraction, the tensor field $\cM$ can be expressed by
\begin{equation}
\label{eq:TensorNess}
\cM(x)=\omega(x)\kq_1(x)\kq_1^{\rm T}(x)+\kq_2(x)\kq^{\rm T}_2(x),
\end{equation}
where $\omega(x)=\exp\big(-\alpha P(x)\big)_0$ is a potential  function with $\alpha\in\bR^+$ and  $P:\Omega\to[0,\infty)$ is a vesselness map which has  large values around the vessel centerlines and low values in the background. It can be estimated by the tubularity filters such as~\cite{frangi1998multiscale,law2008three}.
The vector $\kq_1(x)\in\bR^2$ denotes an orientation that a  vessel should have at point $x$ and $\kq_2(x)$ is orthogonal to $\kq_1(x)$.

\begin{figure*}[t]
\centering
\includegraphics[width=8.5cm]{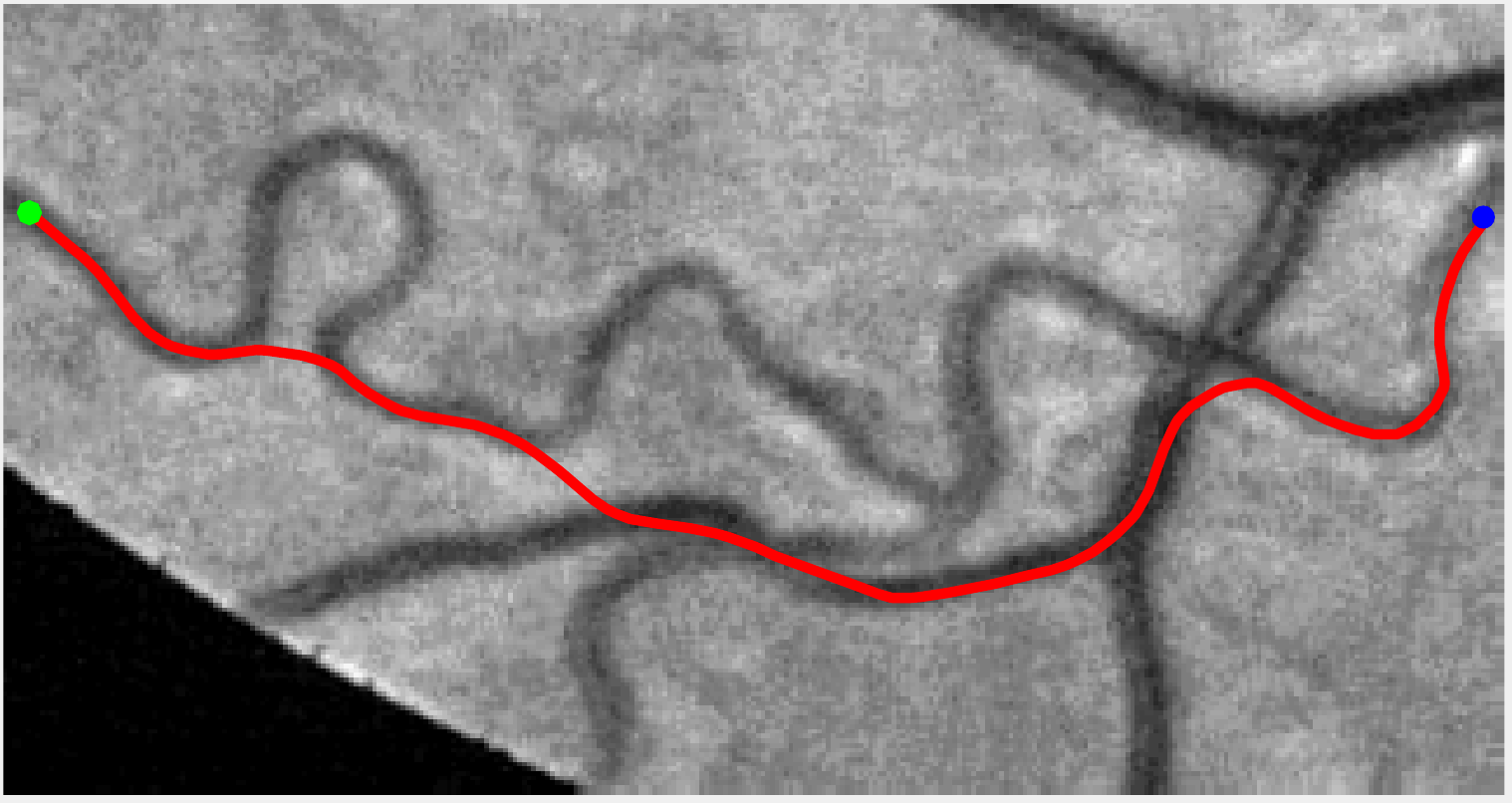}
\includegraphics[width=8.5cm]{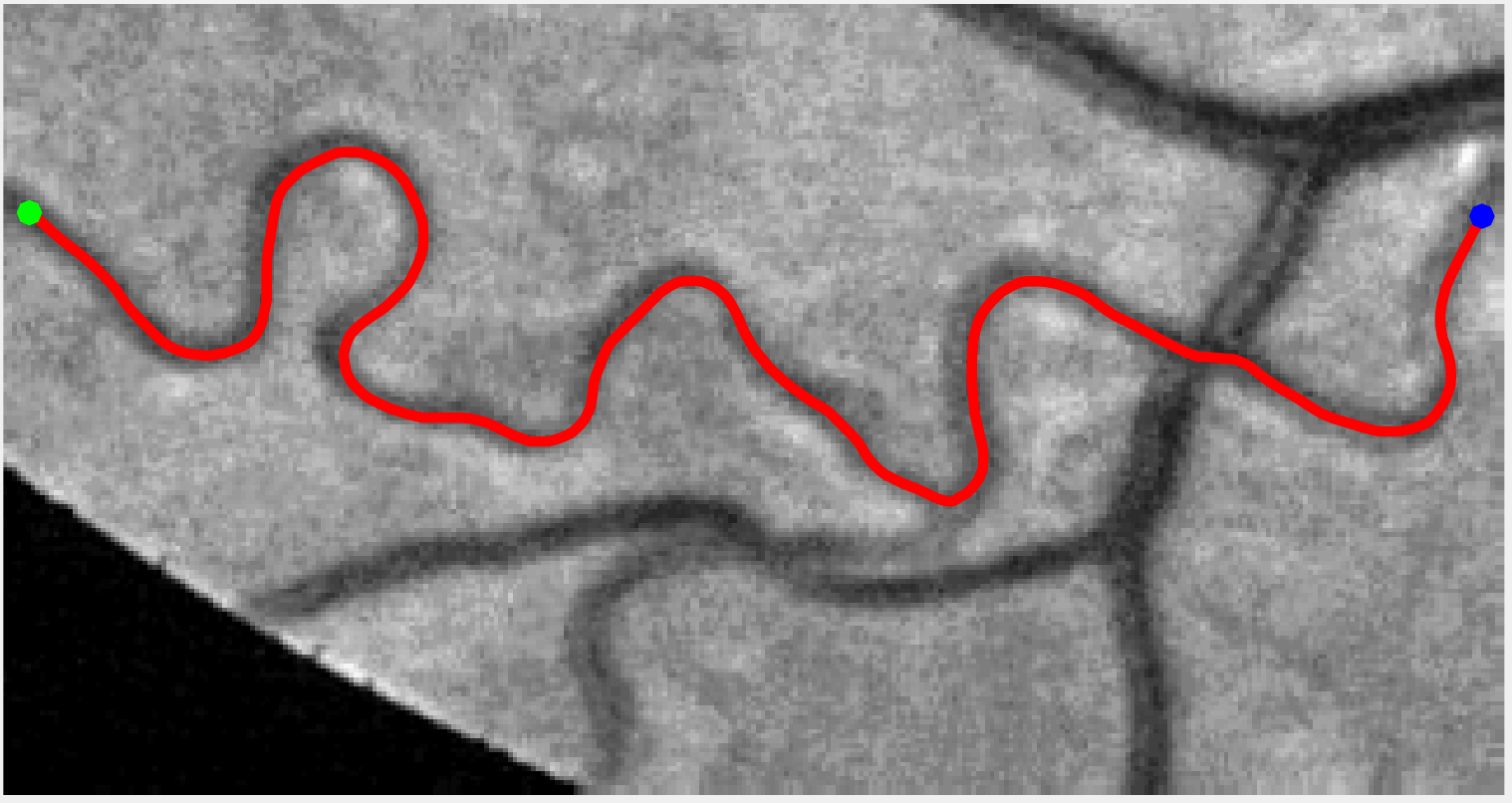}
\caption{\textbf{Left} Result by the curvature-penalized model~\cite{chen2017global}. \textbf{Right} Result from the proposed model after refinement step.}
\label{fig:PS}
\end{figure*}

\section{Coherence-Penalized Geodesic Model}
In this section, we present a new coherence-penalized metric based on a  scalar-valued  smooth  function $\cI:\Omega\to \bI\subset\bR^+_0$. 
\subsection{Geodesic Energy with Coherence-penalized Metric}
Let us consider a norm $\phi_x$  over $\bR^2$ that depends on  a directional derivative operator $\cD_{\fu}$ such that for any point $x\in\Omega$
\begin{equation}
\label{eq:CoherenceNorm}
\phi_x(\fu)	=|\cD_{\fu}\cI(x)|=\left|\lim_{\epsilon\to 0}\frac{\cI(x+\epsilon\vec u)-\cI(x)}{\epsilon}\right|.
\end{equation}
Therefore, one can point out that a small value of $\cD_{\fu}\cI$ implies a low variation of  $\cI$ along the  direction $\fu$ at a point $x$. We consider a geodesic energy of  a curve $\eta\in\Lips([0,1],\Omega)$
\begin{equation}
\label{eq:CoherencePathEnergy}
\cL(\eta)=\int_0^1\sqrt{\|\dot\eta(t)\|_{\cM(\eta(t))}+\beta\,\cH^2(\eta(t),\dot\eta(t))}\,dt,
\end{equation}
where  the tensor field $\cM$ is defined in Eq.~\eqref{eq:TensorNess} and $\beta\in\bR^+_0$. The term $\cH$ depends on the norm~\eqref{eq:CoherenceNorm} such that 
\begin{equation}
\label{eq:CoherenceTerm}
\cH(x,\fu)=\omega(x)	\phi_x^2(\fu)=\omega(x)|\langle  \nabla \cI(x),\fu \rangle|^2,
\end{equation}
where the second equality  is obtained by the reality $\cD_{\fu}\cI(x)=\langle\nabla \cI(x),\fu\rangle$. The potential function is used in~\eqref{eq:TensorNess}.

\subsection{Minimizing $\cL$ by a lifted metric approach}
\label{sec:LiftSolution}
\begin{definition}
We define a space $\bar\Omega=\Omega\times\bI \subset\bR^3$ such that a point $\bar x=(x,\vartheta)\in \bar\Omega$ is a pair comprised of a spatial point $x\in\Omega$ and a position $\vartheta\in\bI$.  
\end{definition}

Let $\tau:[0,\,1]\to\bI$ be a differentiable parametric function which is defined in terms of $\cI$ and a path $\eta\in\Lips([0,1],\Omega)$
\begin{equation}
\label{eq:Constraint}	
\tau=\cI\circ\eta,
\end{equation}
based on which,  one has for any $t\in[0,1]$
\begin{equation}
\label{eq:ParametericExpression}
\dot\tau(t)=\frac{\partial \tau}{\partial t}(t)=\frac{\partial \cI(\eta(t))}{\partial t}=\langle\nabla\cI(\eta(t)),\,\dot\eta(t)\rangle.
\end{equation}
Integrating Eqs.~\eqref{eq:CoherenceTerm} and \eqref{eq:ParametericExpression}, for a curve $\eta$ we have 
\begin{equation}
\label{eq:CoherenceTerm2}
\cH^2(\eta(t),\dot\eta(t))=\omega(\eta(t))	|\dot\tau(t)|^2
\end{equation}
Based on the relation~\eqref{eq:Constraint}, for each curve $\eta\in\Lips([0,1],\Omega)$  one can construct a lifting curve $\gamma\in \Lips([0,1],\bar\Omega)$ such that 
\begin{equation}
\gamma(t):=(\eta(t),\tau(t)),\quad \text{and}\quad \dot\gamma(t)=(\dot\eta(t),\dot\tau(t)).
\end{equation}
 Now we can  reformulate the geodesic energy $\cL$ in Eq.~\eqref{eq:CoherencePathEnergy} by
\begin{align}
\cL(\eta)&=\int_0^1\sqrt{\|\dot\eta(t)\|_{\cM(\eta(t))}+\beta\,\omega(\eta(t))|\dot\tau(t)|^2}\,dt,\nonumber\\
\label{eq:LiftedExpress}	
&=\int_0^1 \cR^\infty(\gamma(t),\dot\gamma(t))~dt=\cL^\infty(\gamma),
\end{align}
where $\cR^\infty:\bar\Omega\times\bR^3\to[0,\infty]$ is a degenerated Riemannian metric. It can be expressed for any $\bar x=(x,r)\in\bar\Omega$ and any vector $\bar\fu\in\bR^3$ 
\begin{equation}
\label{eq:DeRiemannian}	
\cR^{\infty}(\bar x,\bar\fu):=
\begin{cases}
\|\bar\fu\|_{\cT(x,\vartheta)},\quad &\text{if}~\vartheta=\cI(x),\\
\infty,&\text{otherwise},
\end{cases}
\end{equation}
where $\cT:\bar\Omega\to S^+_3$ is a PSD  tensor field
\begin{equation}
\label{eq:LiftedTensor}
\cT(x,\vartheta)=
\begin{pmatrix}
\bar\cM(x,\vartheta),&\0\\
\0,&\beta\bar\omega(x,\vartheta)
\end{pmatrix},
\end{equation}
with $\bar\cM(x,\vartheta):=\cM(x)$ and $\bar\omega(x,\vartheta):=\omega(x)$.

The goal is to minimize $\cL^\infty$ in Eq.~\eqref{eq:LiftedExpress} by solving the respective  Eikonal PDE. However, the metric $\cR^\infty$ is too singular for finding the  numerical solutions. We introduce a new Riemannian metric $\cR^\lambda:\bar\Omega\times\bR^3\to[0,\infty]$ based on a penalization parameter $\lambda\gg0$ for any $\bar x=(x,\vartheta)$ and any $\bar\fu\in\bR^3$
\begin{equation}
\label{eq:ApproximatedMetric}
\cR^{\lambda}(\bar x,\vec u):=\kc_\lambda\big(\vartheta,\cI(x)\big)\,\|\vec u\|_{\cT(\x)},
\end{equation}
where $\kc_\lambda:\bI\times\bI\to\bR^+_0$ is a distance map  defined by
\begin{equation}
\label{eq:DistanceFunction}
\kc_\lambda(a,b)=\exp(\lambda\,|a-b|^p), \quad \forall a,~b\in\bI,
\end{equation}
where we set $p=1$ through this paper. 

For a given source point $\bar s\in\bar\Omega$, the  minimal action map $\cU^\lambda_{\bar s}$ associated to the metric $\cR^\lambda$  can be estimated efficiently by the anisotropic  fast marching method~\cite{mirebeau2014anisotropic}. The geodesic $\bar\cC_{\bar{\s},\bar{\x}}$ linking the source point $\bar s$ to a point $\bar x$ can be computed  via the solution to the gradient descent ODE~\eqref{eq:ODE} on $\cU_{\bar s}^\lambda$.

\noindent\textbf{Minimal paths refinement processing.}
In practice, sometimes the resulting geodesics $\bar\cC_{\bar\s,\bar\x}$ by the metric $\cR^\lambda$ do not exactly follow  the real centerlines because of the local gray level inhomogeneities.  Here we use the region-constrained minimal path model proposed in~\cite{chen2015piecewise} to  refine these geodesics. 

\noindent\textbf{Compare to existing minimal path models.}
The Riemannian metrics used in~\cite{cohen1997global,benmansour2011tubular,pechaud2009extraction} are based on the local \emph{pointwise} information. The curvature-penalized metric~\cite{chen2017global,bekkers2015pde} and the proposed coherence-penalized metric are able to consider more constraints, i.e.,  the rigidity for~\cite{chen2017global,bekkers2015pde}   and feature coherence  for our metric. These constraints are beneficial  to the respective geodesics to reduce the risk of short branches combination problem. Compared to the curvature-penalized metric,  our method can be more flexible since the feature map $\cI$ can be produced dependently on the task. In retinal imaging,  veins and arteries are distinguishable in terms of gray levels or  vesselness values,  satisfying the formulation of the proposed model. Especially for vessels with strong tortuosity, the curvature-penalized metric， which favours a smooth curve,  fails to catch the expected vessels as shown in the left column of  Fig.~\ref{fig:PS}. From the right column of Fig.~\ref{fig:PS}, one can see that our model can obtain a good result.

\begin{figure*}[t]
\centering
\includegraphics[width=4.2cm]{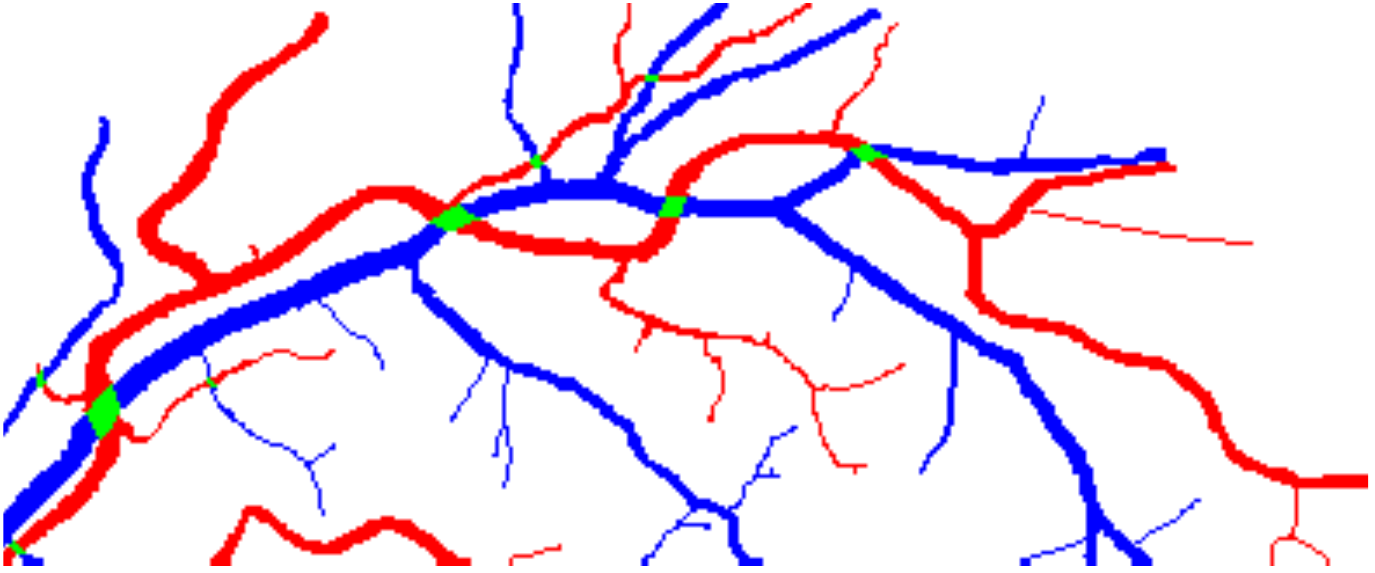}\,\includegraphics[width=4.2cm]{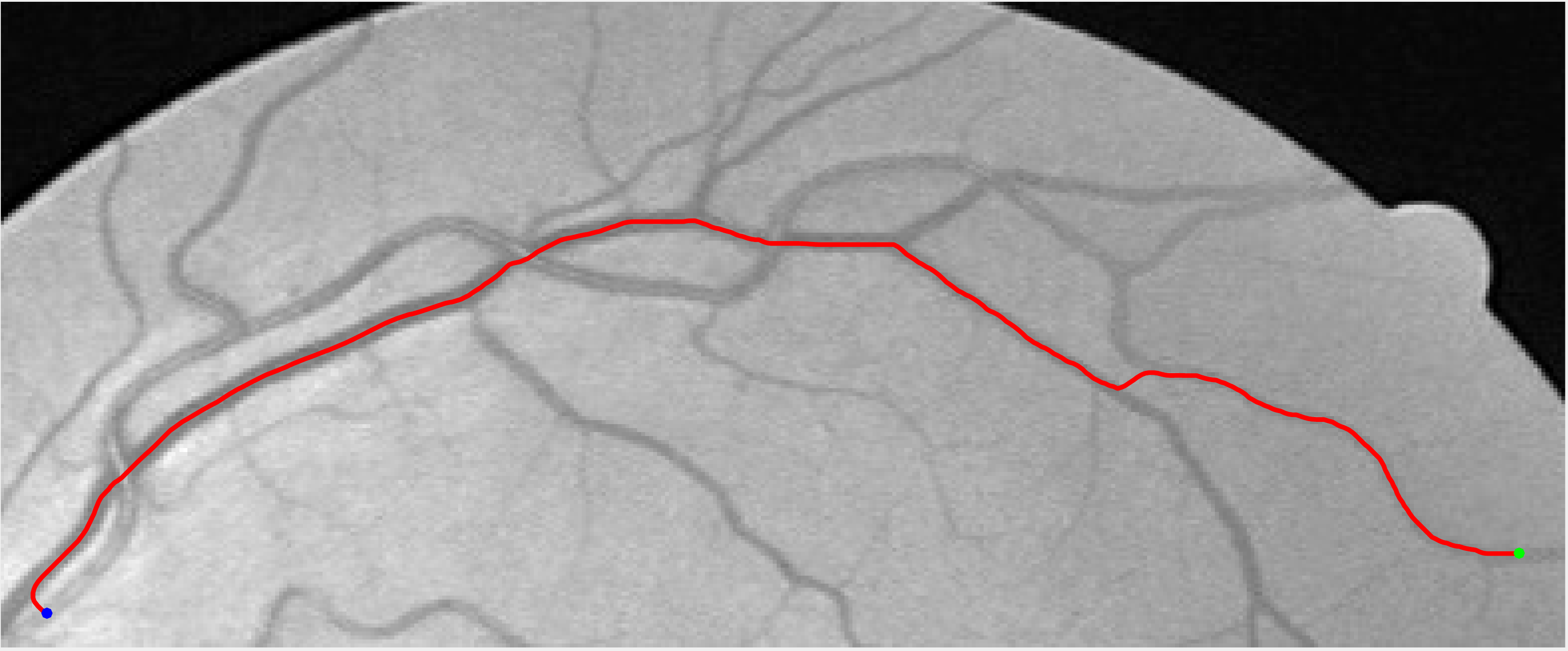}\,\includegraphics[width=4.2cm]{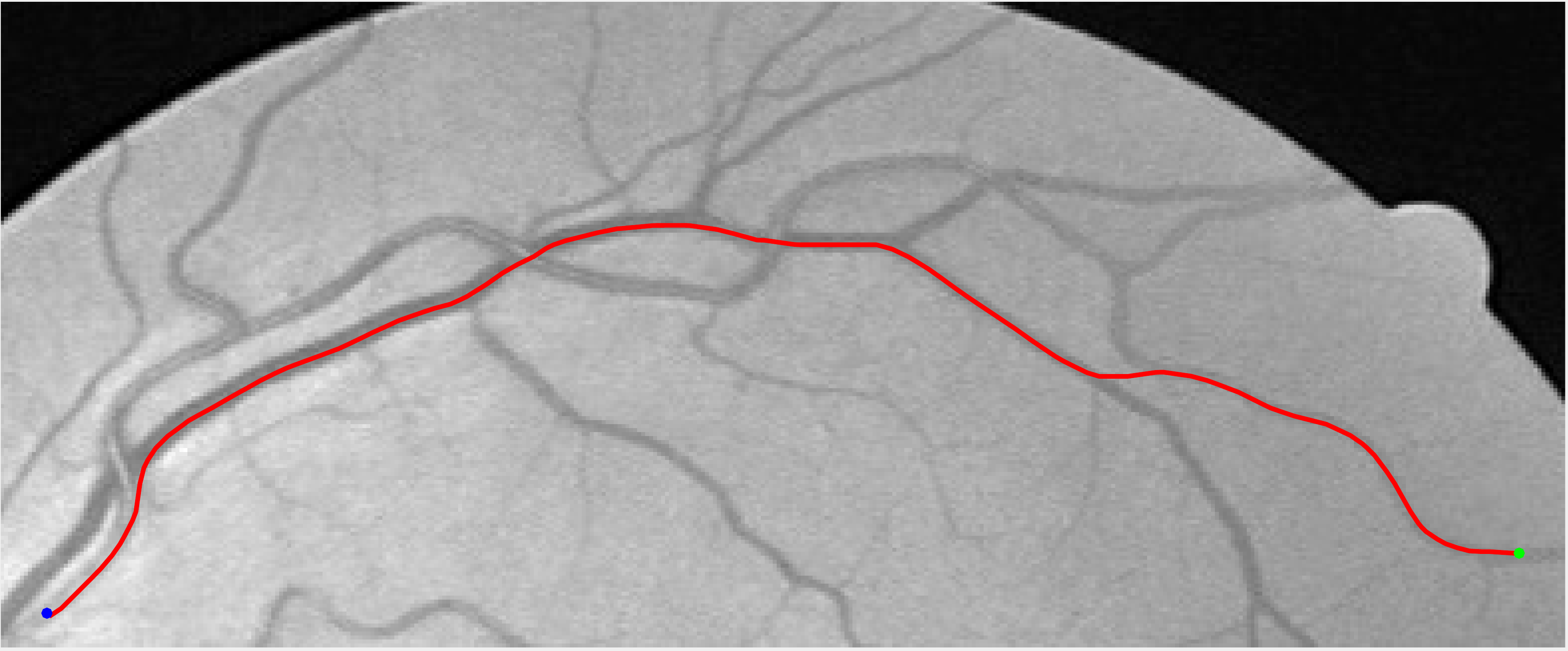}\,\includegraphics[width=4.2cm]{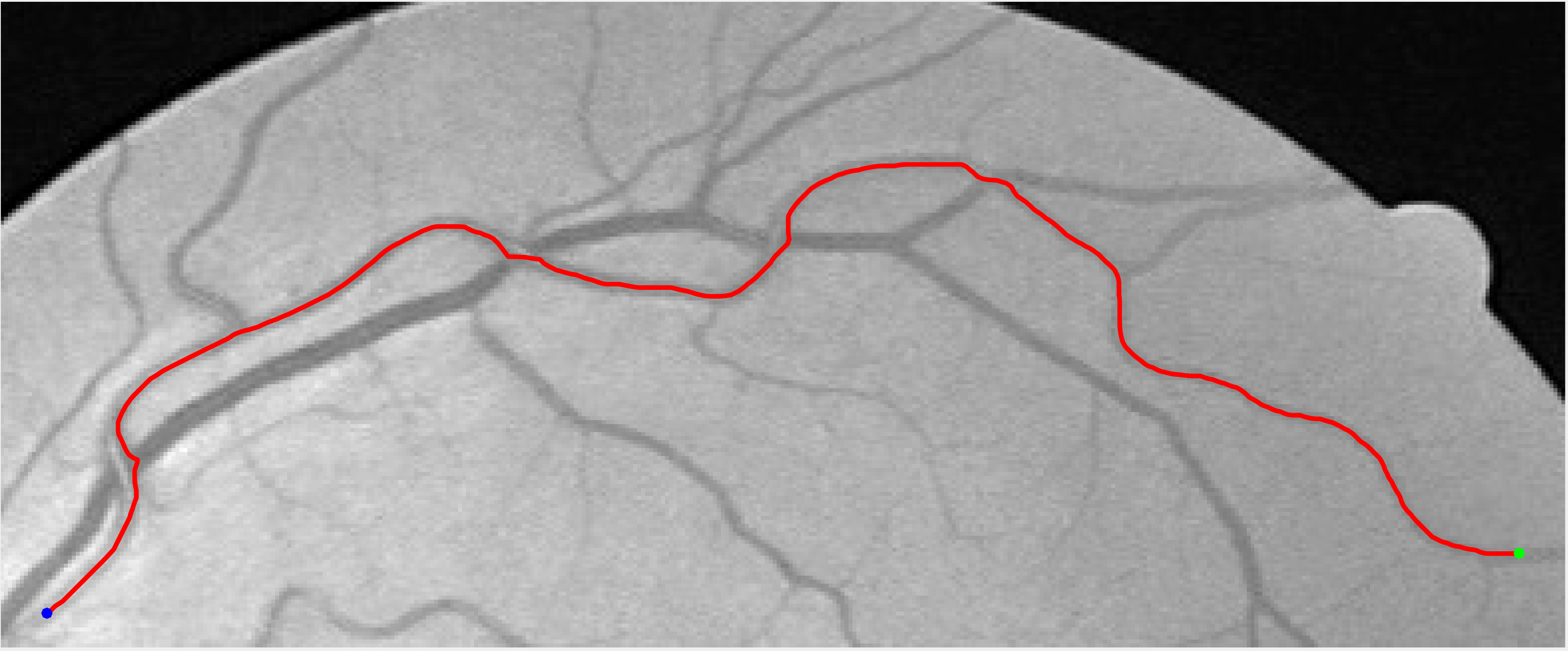}\\
\includegraphics[width=4.2cm]{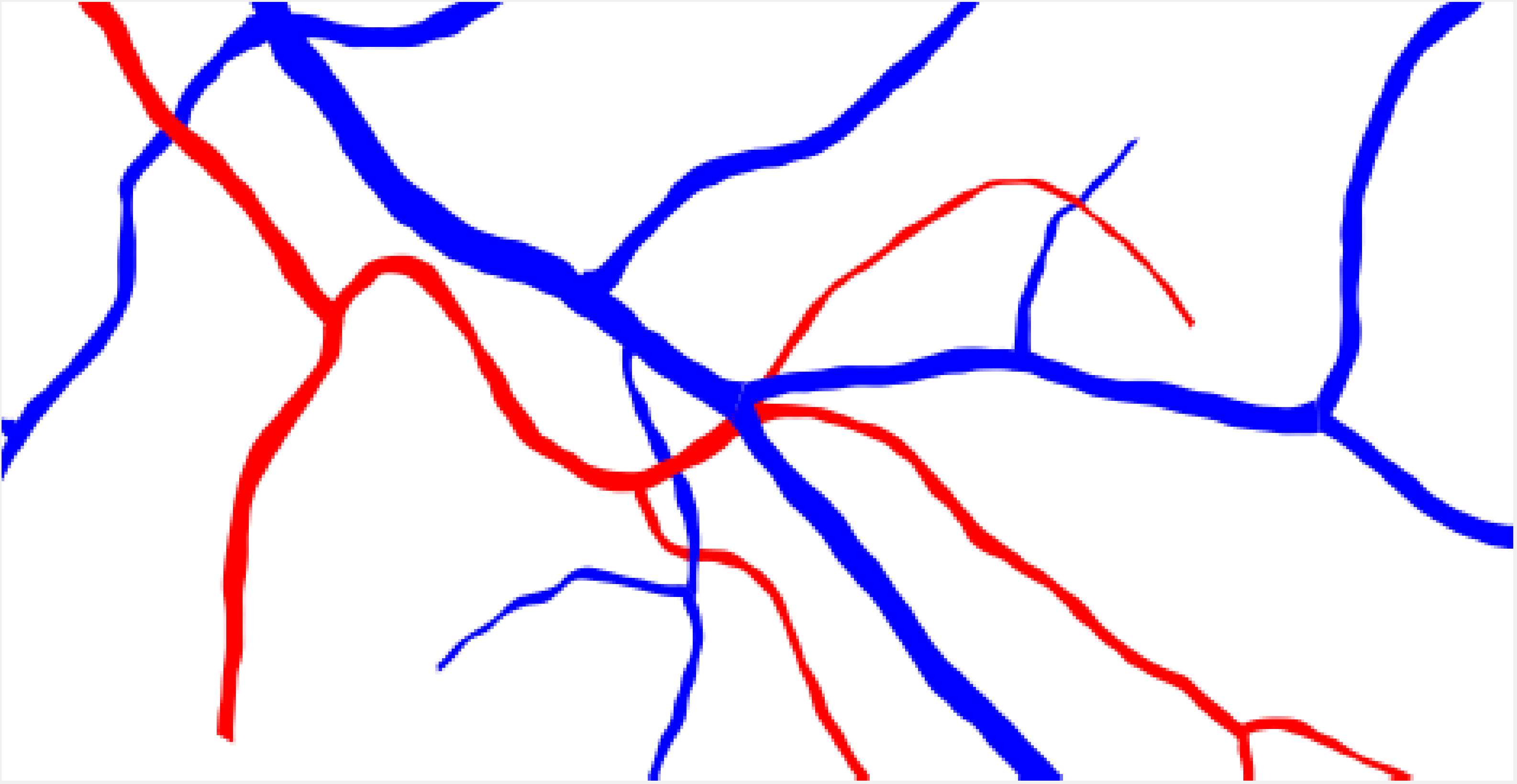}\,\includegraphics[width=4.2cm]{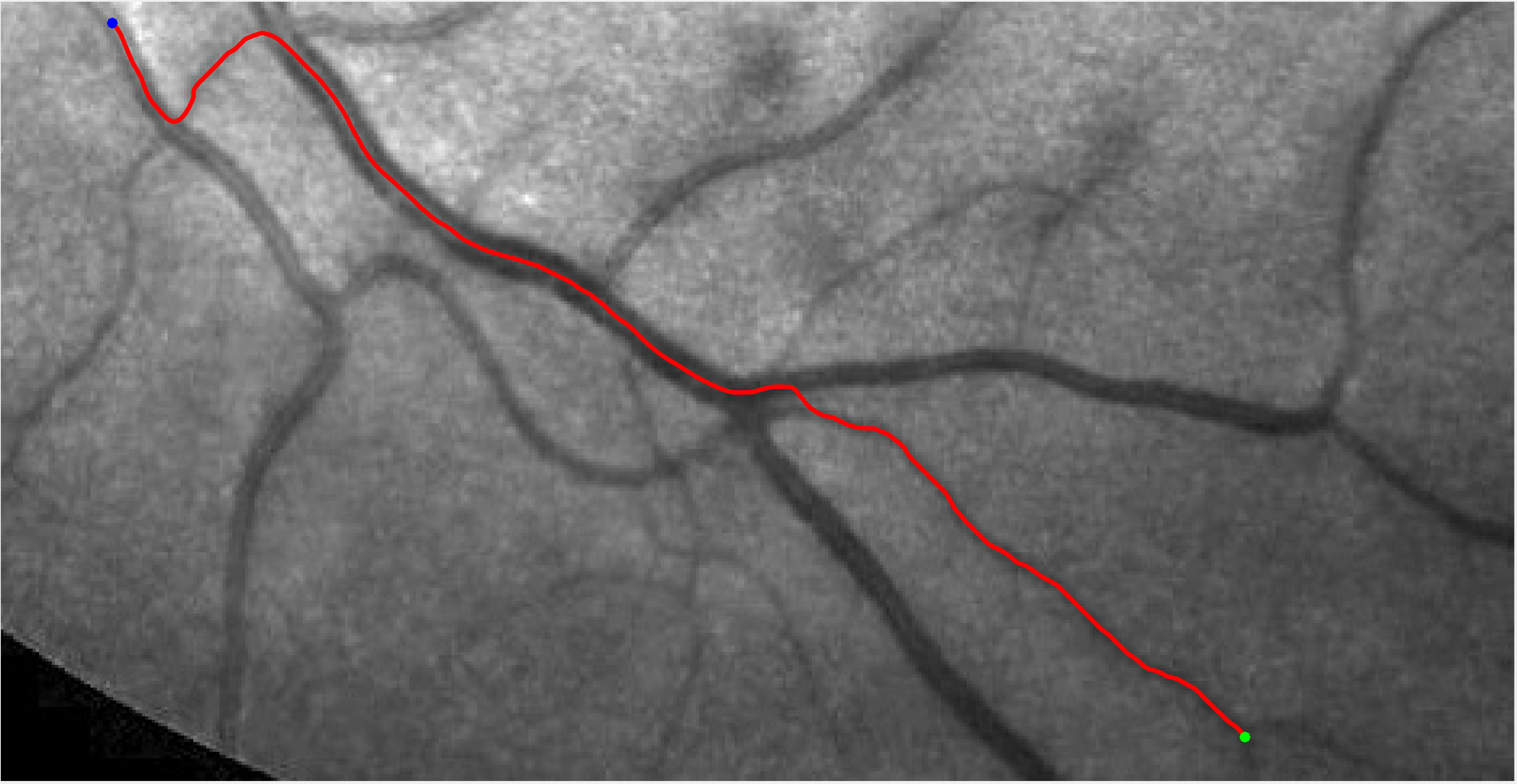}\,\includegraphics[width=4.2cm]{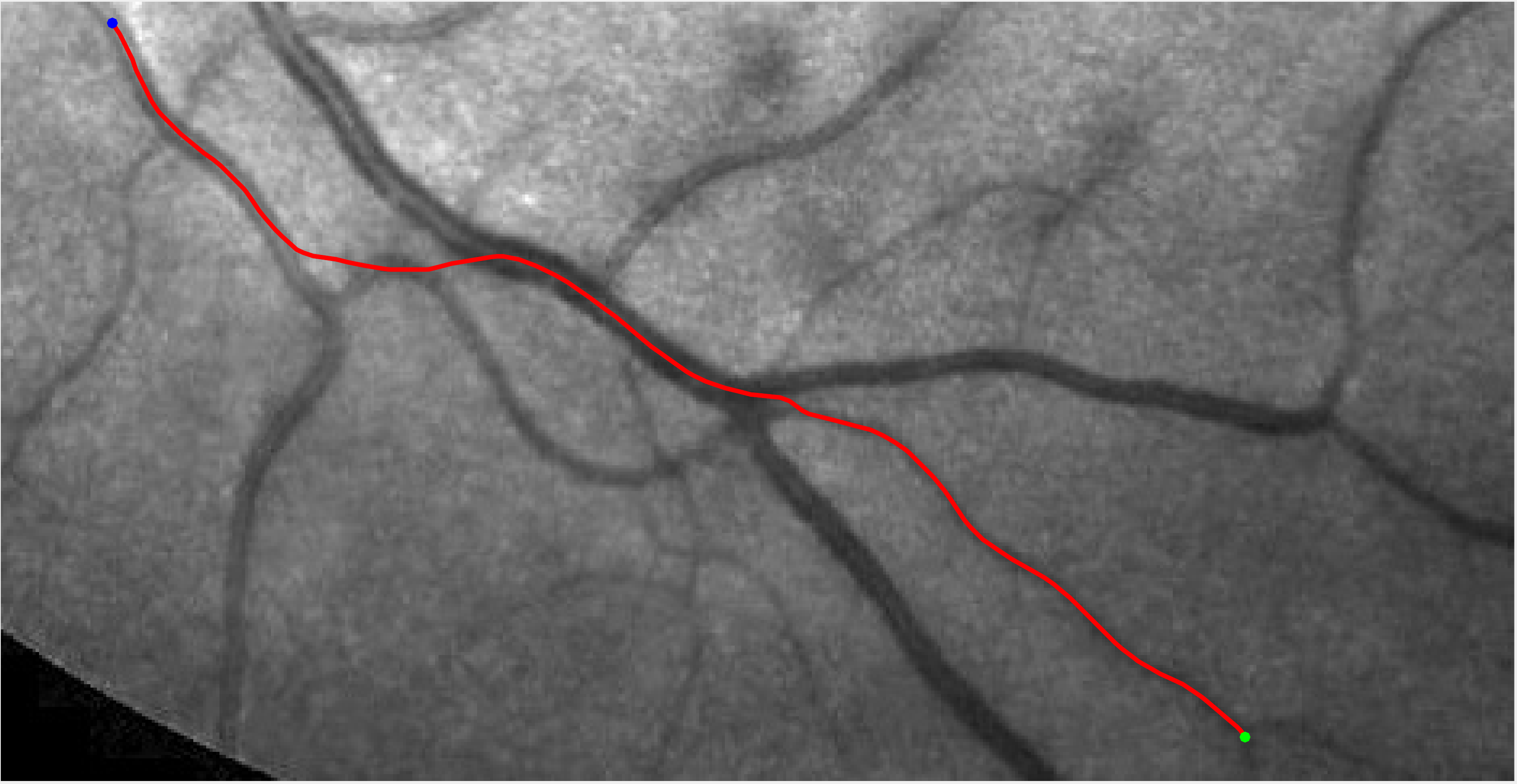}\,\includegraphics[width=4.2cm]{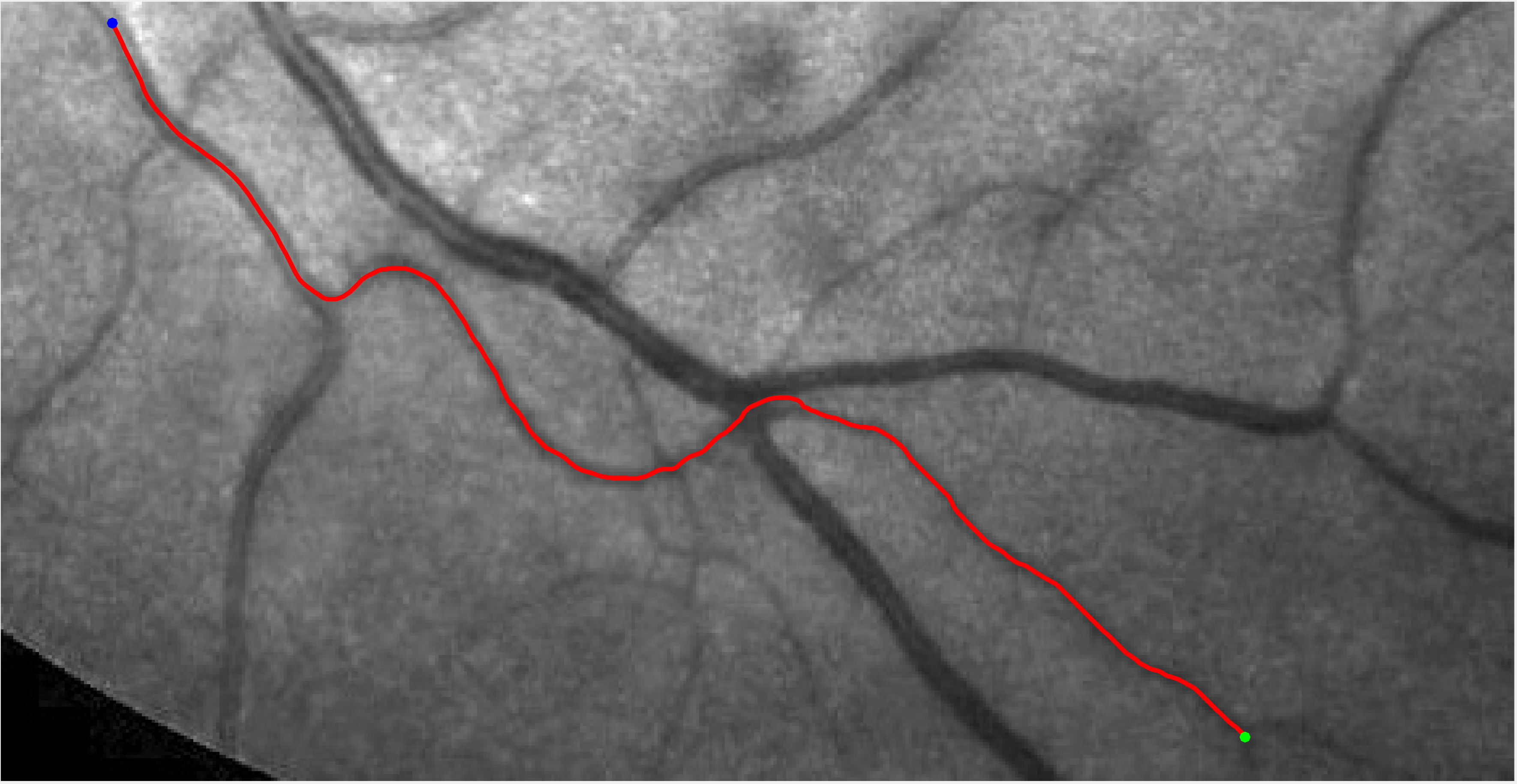}\\
\includegraphics[width=4.2cm]{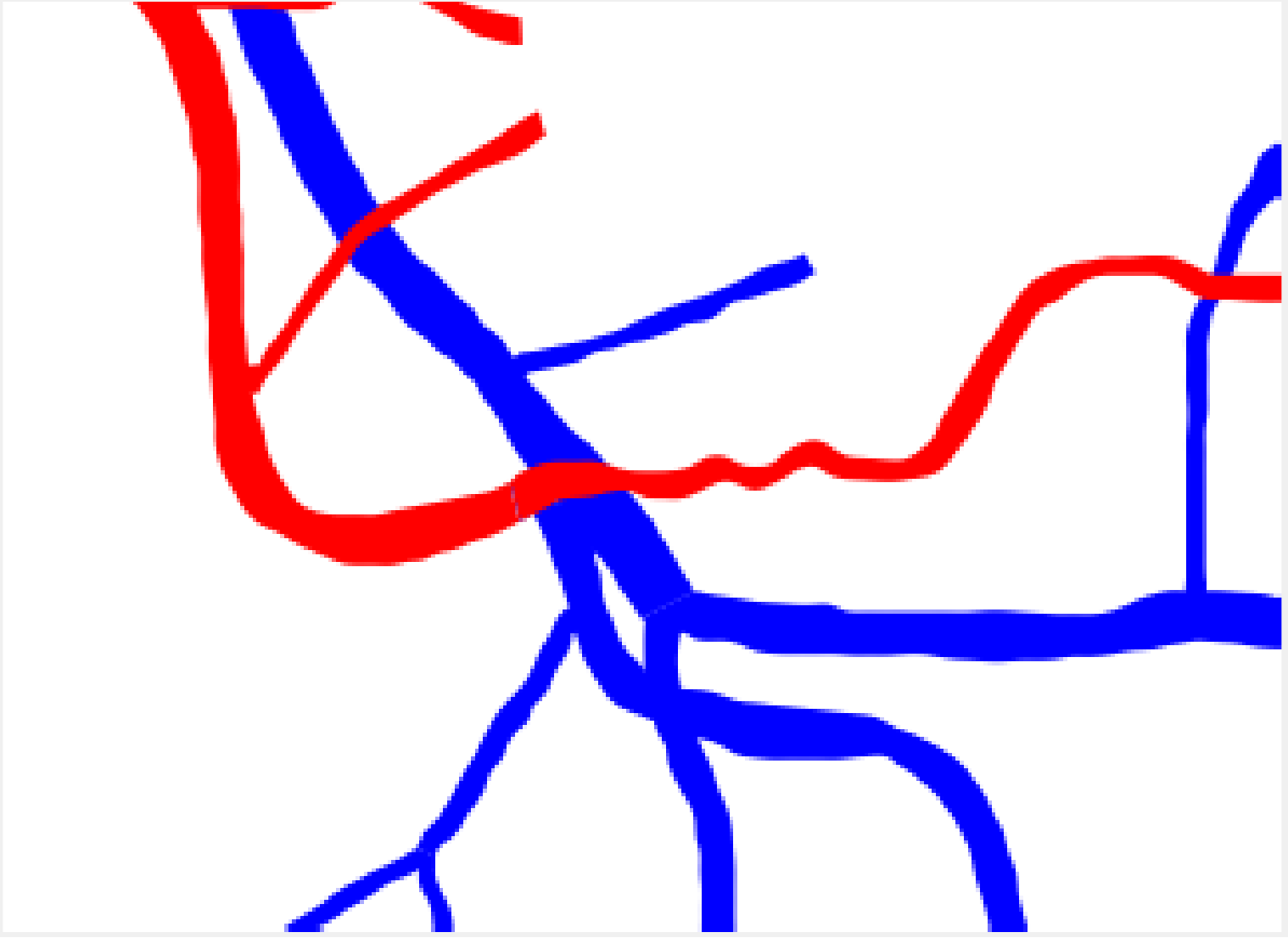}\,\includegraphics[width=4.2cm]{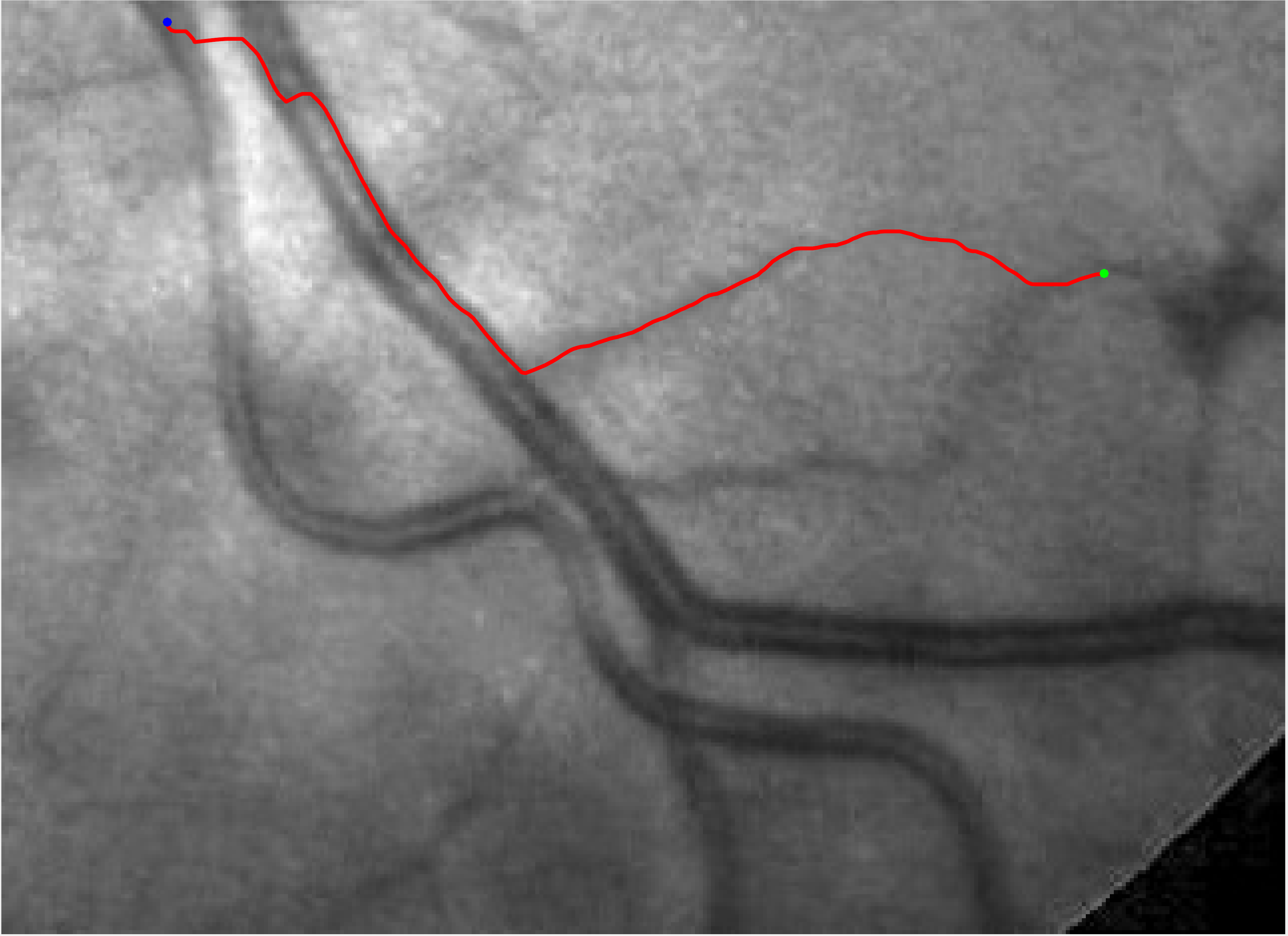}\,\includegraphics[width=4.2cm]{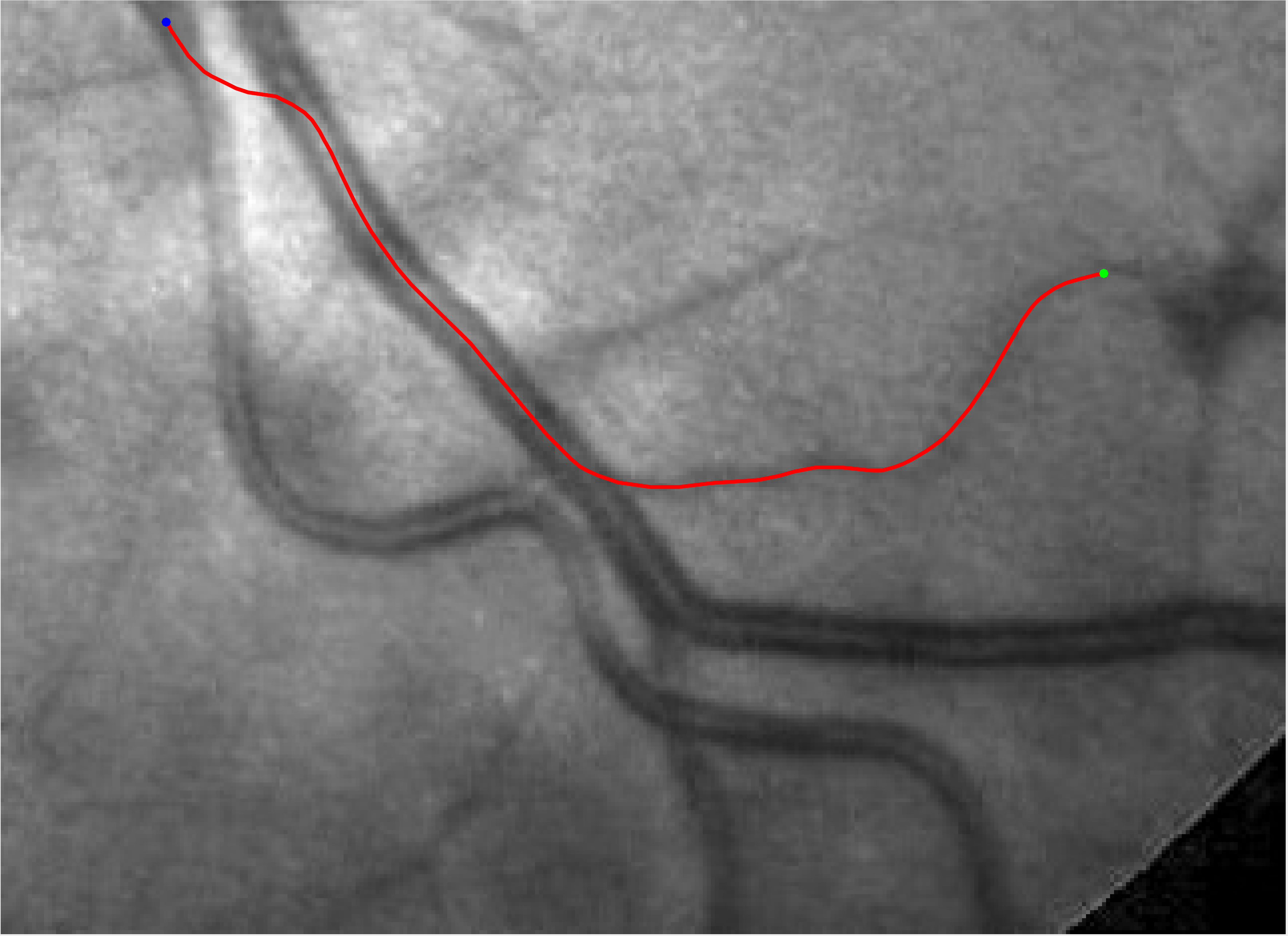}\,\includegraphics[width=4.2cm]{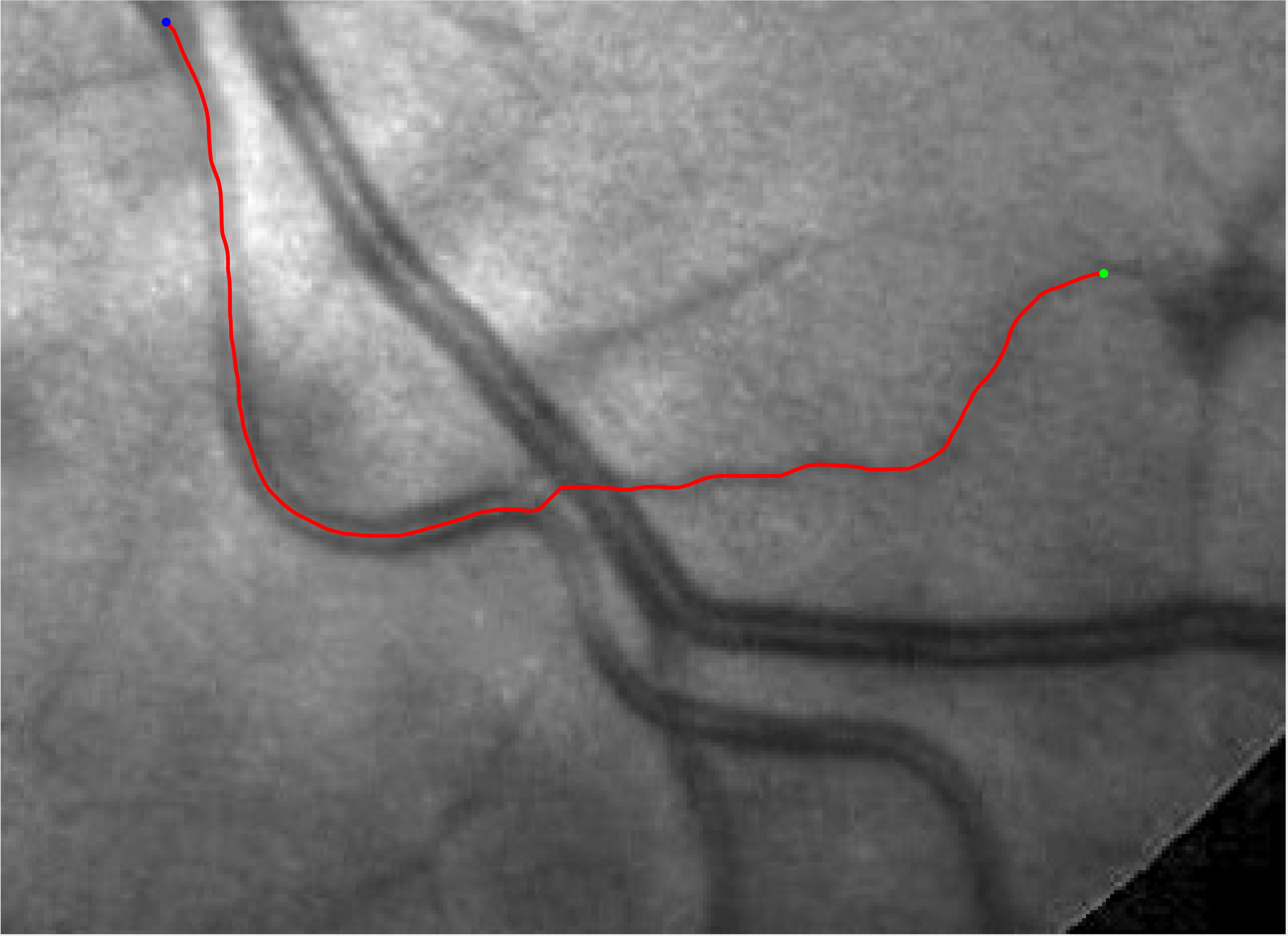}
\caption{\textbf{Column1} shows the AV groundtruth. \textbf{Columns 2-4} are paths from the ArR, CuP and the proposed models respectively.}
\label{fig:examples}	
\end{figure*}

\section{Experimental Results}
\label{sec:Experiments}
\noindent \textbf{Computation of the local vessel geometry.}
There exist a series of methods~\cite{frangi1998multiscale,law2008three} for local vessel geometry computation, among which we choose the optimally oriented flux filter (OOF)~\cite{law2008three} as our feature extractor. The multi-scale outputs of the OOF can be  expressed as $\mathbf Q(x,r)=\left(\frac{1}{r}(\partial_{ij}G_\sigma)_{ij}\ast\mathbbm{1}_r\ast I\right)(x)$, where  $\mathbbm 1_r$ is a step function centred at point  $x$ with scale $r$ and $(\partial_{ij}G_\sigma)_{ij}$ is the Hessian matrix of  the Gaussian kernel $G_\sigma$ with variance $\sigma$.
Generally, we can assume that the gray levels inside the vessels are lower than background and $\xi_1\leq \xi_2$. By decomposing $\mathbf Q$ by its eigenvalues $\hat \xi_n$ and eigenvectors $\hat {\kq}_n$, we can define an optimal scale map  $\cS(x)=	\arg\min_r\hat{\xi}_2(x,r)$.
The vesselness map $P$ can be computed by $P(x)=\max\{0,~\xi_2(x,\cS(x))\}$. The eigenvectors $\kq_n$ of $\cM$ in Eq.~\eqref{eq:TensorNess} can be defined as $\kq_n(x)=\hat {\kq}_n(x,\cS(x))$. 
  The feature map $\cI$ is constructed via a smoothed  vesselness map $\tilde P:=F\ast P$ where $F$ is a mean filter or a Gaussian filter. The interval $\bI=[0, \|\cI\|_\infty]$ is discretized evenly  to  $120$ levels in all the experiments.
 
\noindent \textbf{Validation.}
We validate our minimal path model  on  respective 54 and 30 patches obtained from the DRIVE~\cite{staal2004ridge,hu2013automated} and the IOSTAR~\cite{zhang2016robust} datasets with AV groundtruth. Each artery involved in these parches locates near a vein or crossing it at least once. Our goal is to extract the artery between two given points.
In order to get the quantitative evaluation, we first convert each  continuous spatial path $\cC$ to an 4-connected digital path $\Gamma\subset \Omega$ which is considered as a pixel collection. We denote by $\chi=\{x\in\Gamma\cap A\}$ the collection of digital path pixels inside the artery groundtruth map $A$. Thus, a measure $\Theta$ can be simply defined as $\Theta=|\chi|/|\Gamma|$, where $|\chi|$ and $|\Gamma|$ mean the respective number of elements involved in $\chi$ and $\Gamma$.
We compare our model to four existing minimal path models: the  isotropic Riemannian (IR)  model~\cite{cohen1997global}, the anisotropic radius-lifted  Riemannian (ArR) model~\cite{benmansour2011tubular}, the isotropic orientation-lifted Riemannian (IoR) model~\cite{pechaud2009extraction} and the curvature-penalized (CuP) model~\cite{chen2017global}. The construction of these metrics are based on the OOF outputs~\cite{law2008three}.  Note that a centerline-based potential is chosen so that  we remove the radius dimension of \cite{pechaud2009extraction}  to reduce computation complexity. The results in terms of the $\Theta$ score are presented in Table~\ref{tab:Result}, including the average (Avg.), maximum (Max.),  minimum (Min.) and standard deviation (Std.) values. In both DRIVE and IOSTAR datasets, our method can achieve the best performances thanks to the coherence penalization. Note that in Table~\ref{tab:Result},  we evaluate our method by using the refined paths instead of using the original coherence-penalized minimal paths.
For  comparisons in visualization, we show the minimal paths from the ArR metric , the CuP metric and the proposed coherence-penalized metric on three retinal patches as shown in Fig.~\ref{fig:examples}. The targeted artery vessels which cross veins at least once are labeled by red color in column 1. The paths shown in  column 4 from the proposed metric are results after refinement. One can claim that our method indeed can catch expected arteries while other metrics fall into the traps of short branches combination.

\begin{table}[!t]
\centering
\caption{Quantitative comparisons of  different metrics.}
\label{tab:Result}
\setlength{\tabcolsep}{5pt}
\renewcommand{\arraystretch}{1}
\begin{tabular}{c ||l c c c c c}
\shline
\multicolumn{2}{l}{}$\Theta~\quad$ &IR &ArR  & IoR &CuP  & Proposed \\ 
\hline
\multirow{4}{*}{DRIVE}  & Avg.  & 0.39     & 0.39   & 0.34   & 0.65   & \textbf{0.98}\\
                                    & Max. &  1.0      & 1.0      & 1.0    & 1.0     & 1.0    \\
                                    & Min.  &  0.03    & 0.02   & 0.02   & 0.13   &\textbf{0.83}  \\
                                    & Std.  &  0.22     & 0.28   &0.28    & 0.28   &\textbf{0.04} \\
\hline   
\multirow{4}{*}{IOSTAR}& Avg. & 0.44    & 0.48    & 0.48   & 0.70       &\textbf{0.90}\\
                                    & Max.   & 0.98 &0.98     & 0.97   & 0.97      &\textbf{0.99}\\
                                    & Min.   & 0.02  & 0.03      & 0.03   & 0.06    &\textbf{0.63}\\
                                    & Std.   &  0.30  & 0.32      & 0.32   & 0.30        &\textbf{0.08}\\
\shline
\end{tabular}
\end{table}

\section{Conclusion}
In this paper, we propose a new metric invoking a coherence penalty  on the variation of  an  external feature map along the geodesics  for minimally interactive retinal vessel segmentation. We interpret the geodesic energy with a coherence  penalty through a coherence-penalized metric.
 In order to compute the associated geodesics by fast marching method,  we use an approximation of such a coherence-penalized metric.   Experimental results show that our model indeed obtain promising results in retinal vessel segmentation.

\bibliographystyle{IEEEbib}
\bibliography{isbi2018}

\begin{thebibliography}{10}

\bibitem{cohen1997global}
L.~D. Cohen and Ron Kimmel,
\newblock ``{Global minimum for active contour models: A minimal path
  approach},''
\newblock {\em IJCV}, vol. 24, no. 1, pp. 57--78, 1997.

\bibitem{sethian1999fast}
James~A Sethian,
\newblock ``Fast marching methods,''
\newblock {\em SIAM review}, vol. 41, no. 2, pp. 199--235, 1999.

\bibitem{mirebeau2014anisotropic}
J-M Mirebeau,
\newblock ``Anisotropic fast-marching on cartesian grids using lattice basis
  reduction,''
\newblock {\em SINUM}, vol. 52, no. 4, pp. 1573--1599, 2014.

\bibitem{benmansour2011tubular}
F.~Benmansour and L.~D. Cohen,
\newblock ``Tubular structure segmentation based on minimal path method and
  anisotropic enhancement,''
\newblock {\em IJCV}, vol. 92, no. 2, pp. 192--210, 2011.

\bibitem{pechaud2009extraction}
M.~P{\'e}chaud, R.~Keriven, and G.~Peyr{\'e},
\newblock ``Extraction of tubular structures over an orientation domain,''
\newblock in {\em CVPR}, 2009, pp. 336--342.

\bibitem{bekkers2015pde}
E.~J Bekkers et~al.,
\newblock ``{A PDE approach to data-driven sub-riemannian geodesics in SE
  (2)},''
\newblock {\em SIIMS}, vol. 8, no. 4, pp. 2740--2770, 2015.

\bibitem{chen2017global}
D.~Chen, J-M Mirebeau, and L.D Cohen,
\newblock ``{Global minimum for a Finsler elastica minimal path approach},''
\newblock {\em IJCV}, vol. 122, no. 3, pp. 458--483, 2017.

\bibitem{liao2017progressive}
W.~Liao et~al.,
\newblock ``{Progressive minimal path method for segmentation of 2D and 3D line
  structures},''
\newblock {\em TPAMI}, 2017.

\bibitem{wang2013interactive}
Lu~Wang, Vinutha Kallem, et~al.,
\newblock ``Interactive retinal vessel extraction by integrating vessel tracing
  and graph search,''
\newblock in {\em MICCAI}, 2013, pp. 567--574.

\bibitem{ulen2015shortest}
J.~Ul{\'e}n et~al.,
\newblock ``Shortest paths with higher-order regularization,''
\newblock {\em TPAMI}, vol. 37, no. 12, pp. 2588--2600, 2015.

\bibitem{frangi1998multiscale}
A.~F Frangi, W.~J Niessen, K.~L Vincken, and M.~A Viergever,
\newblock ``Multiscale vessel enhancement filtering,''
\newblock in {\em MICCAI}, 1998, pp. 130--137.

\bibitem{law2008three}
M.~WK Law and A.~CS Chung,
\newblock ``{Three dimensional curvilinear structure detection using optimally
  oriented flux},''
\newblock in {\em ECCV}, 2008, pp. 368--382.

\bibitem{chen2015piecewise}
D.~Chen and L.~D Cohen,
\newblock ``{Piecewise geodesics for vessel centerline extraction and boundary
  delineation with application to retina segmentation},''
\newblock in {\em SSVM}, 2015, pp. 270--281.

\bibitem{staal2004ridge}
J.~Staal et~al.,
\newblock ``Ridge-based vessel segmentation in color images of the retina,''
\newblock {\em TMI}, vol. 23, no. 4, pp. 501--509, 2004.

\bibitem{hu2013automated}
Q.~Hu et~al.,
\newblock ``Automated separation of binary overlapping trees in low-contrast
  color retinal images,''
\newblock in {\em MICCAI}, 2013, pp. 436--443.

\bibitem{zhang2016robust}
J.~Zhang et~al.,
\newblock ``Robust retinal vessel segmentation via locally adaptive derivative
  frames in orientation scores,''
\newblock {\em TMI}, vol. 35, no. 12, pp. 2631--2644, 2016.

\end{thebibliography}

\end{document}